\documentclass[onecolumn,showpacs,showkeys,superscriptaddress]{revtex4}

\usepackage{amsmath}
\usepackage{bbold}
\usepackage{amsfonts}
\usepackage{amssymb}
\usepackage{pbsi}
\usepackage[T1]{fontenc}
\usepackage{xcolor}
\usepackage{graphicx}
\usepackage{subfig}
\usepackage{float}
\usepackage{color}
\usepackage{hyperref}
\hypersetup{colorlinks = true,
	linkcolor = red,
	anchorcolor = blue,
	citecolor = blue,
	filecolor = blue,
	urlcolor = blue}

\def\openone{\leavevmode\hbox{\small1\kern-3.8pt\normalsize1}}
\def\N{\leavevmode\hbox{ Z \kern-8 pt\normalsize{Z}}}
\def\openone{\leavevmode\hbox{\small1\kern-3.8pt\normalsize1}}
\def\openJ{\leavevmode\hbox{J \kern-9.5pt\normalsize J}}
\def\openS{\leavevmode\hbox{ S \kern-9.3pt\normalsize S}}

\newcommand{\bb}{\begin{equation}}
\newcommand{\ee}{\end{equation}}
\newcommand{\eqb}{\begin{eqnarray}}
\newcommand{\eqf}{\end{eqnarray}}

\begin{document}

\title{Supersymmetric behavior of polarized electromagnetic waves in anisotropic media}

\author{Felipe A. Asenjo}
\email{felipe.asenjo@uai.cl}
\affiliation{Facultad de Ingenier\'ia y Ciencias,
Universidad Adolfo Ib\'a\~nez, Santiago 7491169, Chile.}
\author{Sergio A. Hojman}
\email{sergio.hojman@uai.cl}
\affiliation{Departamento de Ciencias, Facultad de Artes Liberales,
Universidad Adolfo Ib\'a\~nez, Santiago 7491169, Chile.}
\affiliation{Departamento de F\'{\i}sica, Facultad de Ciencias, Universidad de Chile,
Santiago 7800003, Chile.}
\affiliation{Centro de Recursos Educativos Avanzados, CREA, Santiago 7500018, Chile.}
\author{Braulio M. Villegas-Mart\'inez}
\email{bvillegas@inaoep.mx}
\affiliation{Instituto Nacional de Astrof\'isica, \'Optica y Electr\'onica. Calle Luis Enrique Erro No. 1,
Santa Mar\'ia Tonantzintla, Puebla 72840, Mexico.}
\author{H\'ector M. Moya-Cessa}
\email{hmmc@inaoep.mx}
\affiliation{Instituto Nacional de Astrof\'isica, \'Optica y Electr\'onica. Calle Luis Enrique Erro No. 1,
Santa Mar\'ia Tonantzintla, Puebla 72840, Mexico.}
\author{Francisco Soto-Eguibar}
\email{feguibar@inaoep.mx}\affiliation{Instituto Nacional de Astrof\'isica, \'Optica y Electr\'onica. Calle Luis Enrique Erro No. 1,
Santa Mar\'ia Tonantzintla, Puebla 72840, Mexico.}

\begin{abstract}
A medium with specific anisotropic refractive indices can induce a supersymmetric behavior in the propagation of polarized electromagnetic waves, in an analogue fashion to a quantum mechanical system. The polarizations of the wave are the ones which behave as superpartners from each other. For this to happen, the anisotropy of the medium must be transverse to the direction of propagation of the wave, with different refractive indices along the direction of each polarization. These refractive indices must follow a very specific relation in order to trigger the supersymetric response of the electromagnetic wave, each of them with spatial dependence on the longitudinal (propagation) direction of the wave. In this form, in these materials, different polarized light can be used to test supersymmetry in an optical fashion.
\end{abstract}

%\pacs{04.60 Bc, 98.80 Qc, 11.30 Cp}

\maketitle

\section{Introduction}
Supersymetry can be considered as a fundamental idea that is ubiquitous to several fields in physics (see, for example, Refs.~\cite{freund,David,negro,coop1,coop2,crom,jfeng,hall,zach,haber,aiit,qcosmo1,qcosmo2,qcosmo3,hojmanasenjo,carena,simon,cmscol,vladi,rouz,heweet,sunyu,Plyushchay1,Plyushchay2,correaf,Efetov,junker,Correa22,mateosG,nieto22,Plyushchay3,adrianS,fazsah17,fazsusyt, kwolf,MatthiasHeinrich,carlos}, and references therein).
On the other hand, electromagnetism is one of the most powerful physical theories, that has the larger range of application in our world. Therefore, one can ponder on the possibility of taking advantage of the experimental manipulation of electromagnetic waves in order to devise a form to construct or to measure supersymmetric features. This enterprise has already been initiated by describing light using Helmoltz equation; the spirit of such idea is to focus in the scalar properties of an unpolarized light ray in order to construct its supersymmetric properties  in space-domain  \cite{kwolf,MatthiasHeinrich} or in time-domain \cite{carlos}. In both cases, supersymemtric behavior is achieved by defining two different refractive indices that allow two light-rays to behave as supersymmetric partners. This idea has been shown to be successful, at least theoretically, in properly describing light-rays superpartners depending on what kind of material they propagate. However, this concept can be taken further, in order not just to describe light in the scalar (eikonal) limit, but considering its wave characteristics, such as polarization, in a way to define and construct supersymmetric properties. 

The purpose of this work is to show that transverse polarization components of electromagnetic waves can follow a supersymmetric structure for some specific anisotropic media. Specific anisotropic medium induces the supersymmetry on the electromagnetic wave propagation in a way that each polarization component of a transverse electromagnetic wave becomes the superpartner of the other. The medium to be analyzed in this work is in the space-domain (although the time-domain extension is straightforward), and in order to display  supersymmetric behavior, the anisotropy of this medium, represented by different refractive indices,
must be transversal to the wave propagation, i.e., each polarization (transversal to the longitudinal direction of propagation) interact with a different refractive index. However, both refractive indices must depend on the longitudinal direction of propagation.
Assuming that the electromagnetic wave propagates in a longitudinal direction, say $z$-direction, each transverse polarization to the longitudinal direction (we use $+$ and $-$ to identify the two polarizations) must propagate in an different and specific effective medium with refractive indices $n_+$ and $n_-$, with the specific relations
\begin{align}
  n_+^2+n_-^2&=(1+\zeta)^2\, ,\nonumber\\
  n_+^2-n_-^2&=\frac{2 i c}{\omega}\frac{d\zeta}{dz}\, ,
  \label{relationnpnm}
\end{align}
where $\omega$ is the electromagnetic wave frequency, $c$ is the speed of light, and $\zeta=\zeta(z)$ is an arbitrary complex function. Whenever an electromagnetic wave propagates in an anisotropic transversal medium, with refractive indices fulfilling  relations \eqref{relationnpnm}, the two possible polarizations of the wave  behave as superpartners in a supersymmetric theory. The arbitrary function $\zeta$ must depend on the longitudinal propagation direction (otherwise, the medium becomes isotropic). Thus, for an anisotropic media to be used to construct supersymmetric properties of an electromagnetic wave, in general, their refractive indices must be not constant complex functions.

In the following, we show how the above refractive indices \eqref{relationnpnm} are consequences of Maxwell equations under the imposition of supersymmetry for polarized electromagnetic waves.  Besides, we show some exact and approximated solutions for different refractive indices that allow the supersymmetric behavior.  In the Conclusion section,  we mention
the  possibilities of construct such kind of waves and medium.
   
\section{Supersymmetric electromagnetic polarization partners}

Let us consider general electromagnetic waves in a media described by the Maxwell equations without sources \cite{jackson}
\begin{align}
\nabla\cdot{\bf D}&=0\, ,\quad   &\nabla\cdot{\bf B}&=0 \, ,
\nonumber\\
\nabla\times{\bf E}&=-\frac{\partial{\bf B}}{\partial t}\, ,\quad
&\nabla\times{\bf H}&=\frac{\partial{\bf D}}{\partial t}  \, ,
\label{maxwell1}
\end{align}
where the electric displacement vector is given by ${\bf D}=\epsilon\otimes {\bf E}$, in terms of the electric field ${\bf E}$, and the permittivity tensor $\epsilon$. Similarly, the magnetic field is given by ${\bf B}=\mu\otimes {\bf H}$, in terms of the magnetizing field ${\bf H}$ and the  permeability tensor $\mu$. 

We start by simply describing a non-magnetic anisotropic medium in such form that the off-diagonal terms in the permittivity tensor vanish ($\epsilon_{ij}=0=\mu_{ij}\, , \forall\, i\neq j$), and its diagonal terms are all different from each other ($\epsilon_{11}=\epsilon_x$, $\epsilon_{22}=\epsilon_y$, $\epsilon_{33}=\epsilon_z$); also, let us consider, for simplicity, that permittivity components $\epsilon_i$ do not depend on time. Furthermore, as the medium is non-magnetic, the permeability tensor becomes proportional to the identity, such that its components are of the form $\mu_{ij}=\mu_0 \delta_{ij}$, with $\mu_0$ the vacuum permeability constant and $\delta_{ij}$ the Kronecker delta function. Therefore, under these assumptions, from Eqs.~\eqref{maxwell1}, we find the wave equation for the electric field components
\begin{equation}
      \mu_0\epsilon_i\frac{\partial^2 E_i}{\partial t^2}= \nabla^2 E_i+\partial_i\left(\frac{\partial_j\epsilon_j}{\epsilon_j}E_j\right)\, ,
      \label{waveequGen}
\end{equation}
written for each component of the electric field ($i=x,y,z$), and where the Einstein summation convention is invoked for the last term.

For a transverse wave, with frequency $\omega$, and  form 
\begin{equation}
       {\bf E}(z,t)=\left(E_x(z),E_y(z),0\right)e^{i\omega t}\, ,
\end{equation}
propagating in a media that has only space dependence in  the $z$-direction, Eq.~\eqref{waveequGen} reduces to 
\begin{eqnarray}
      0&=& \frac{d^2 E_\pm}{dz^2}\, +\frac{\omega^2}{c^2} n^2_\pm E_\pm,
      \label{waveequGen2}
\end{eqnarray}
for each polarization component of the electromagnetic wave, that we have chosen to write as $E_x\equiv E_+$ and $E_-\equiv E_y$. Similarly, we have renamed the permittivities as $\epsilon_x\equiv \epsilon_+$ and $\epsilon_y\equiv \epsilon_-$ (here $\epsilon_0$ is the vacuum permittivity constant), and therefore we have defined the corresponding refractive indices $n_\pm(z)=\sqrt{\epsilon_\pm(z)/\epsilon_0}$ for each transversal direction.
   
Supersymmetry between electromagnetic wave polarizations can only occur  when $E_+\neq E_-$, provided that $n_+\neq n_-$ ($\epsilon_+\neq \epsilon_-$). In such case, supersymmetry can be achieved when
Eqs.~\eqref{waveequGen2} can be written  as
\begin{equation}
H_\pm E_\pm=Q_\mp Q_\pm E_\pm=0 \, ,
\label{susy1}
\end{equation}
with the operators \cite{coop2}
\begin{equation}
Q_\pm=\pm\frac{d}{dz}+W \, ,
\end{equation}
as long as that the anisotropic medium has permittivity transversal components given exactly by the form
\begin{equation}
n^2_\pm=\frac{c^2}{\omega^2}\left(\pm\frac{dW}{dz}-W^2 \right)\, .
\label{permipluminus}
\end{equation}
Thus, the $W$ function (with  units of inverse length) plays the key role to obtain the supersymmetry between the polarizations. Variations of $W$ measure the anisotropy of the medium. Hence, supersymmetric behavior of polarized electromagnetic waves is lost when $W$ is constant, implying that in such cases we can chose $W=i \omega/c$, in order to obtain the vacuum refractive index. Then, without loss of generality, we can always write this function as
\begin{equation}
    W(z)=i\frac{\omega}{c}\left[1+\zeta(z)\right]\, ,
    \label{definedw}
\end{equation}
such that the function $\zeta(z)$ measures now the type of anisotropy of the the medium that induces supersymmetric behavior on electromagnetic waves. In this way, previous Eqs.~\eqref{permipluminus} and \eqref{definedw} imply relations \eqref{relationnpnm}.
 
The above description, using the anisotropic refractive indices \eqref{permipluminus}, is a supersymmetric theory where the two polarization of the electromagnetic wave behave as superpartners of each other, forming the whole electromagnetic wave a supersymmetric structure.
In this sense, all the framework for a standard supersymmetric theory in quantum mechanics can be straightforwardly utilized. Its algebra can be  defined by the super-Hamiltonian matrix operator ${\bf H}$, and the super-charge matrix operators ${\bf Q}$, defined as \cite{coop2}
\begin{equation}
{\bf H}=\left(\begin{array}{cc}
   H_+ & 0 \\
    0 & H_-
\end{array}\right),  \, \, 
 {\bf Q}=\left(\begin{array}{cc}
       0 & 0 \\
        Q_+ & 0
\end{array}\right), \, \, 
{\bf Q}^\dag=\left(\begin{array}{cc}
       0 & Q_- \\
        0 & 0
    \end{array}\right), 
\end{equation}
having the closed algebra ${\bf H}=\{{\bf Q},{\bf Q}^\dag\}$, $[{\bf H},{\bf Q}]=0=[{\bf H},{\bf Q}^\dag]$, $\{{\bf Q},{\bf Q}\}=0=\{{\bf Q}^\dag,{\bf Q}^\dag\}$ for the bosonic ${\bf H}$, and fermionic ${\bf Q}$ and ${\bf Q}^\dag$ operators.
 
This supersymmetric behavior determines the form of propagation of the polarized electromagnetic waves, which, in general, can be found through Eqs.~\eqref{susy1}.  The two polarizations can be put in the form
\begin{equation}
E_\pm= E_{0} \exp\left(i\frac{\omega}{c}z+i\frac{\omega}{c}\int \chi_\pm dz \right),
\label{formelectromagneticpolariz}
\end{equation}
where $E_{0}$ is an arbitrary integration constant. Using Eqs.~\eqref{susy1}, \eqref{permipluminus} and \eqref{definedw}, we can readily find that
\begin{equation}
    \chi_-=\zeta\, ,
     \label{eqpolartioemmenos}
\end{equation}
where as $\chi_+$ satisfies the equation
\begin{equation}
    i \frac{d \chi_+}{dz}-\frac{\omega}{c}\left(2 \chi_+ +\chi_+^2\right)+i\frac{d\zeta}{dz}+\frac{\omega}{c}\left(2\zeta+\zeta^2\right)=0\, .
    \label{eqpolartioem}
\end{equation}

 Finally, this specific form of supersymmetric polarized propagation can be characterized by 
 the Stokes parameters \cite{jackson,hecht,goldstein}. In the  current general case, normalized Stokes parameters are given by 
\begin{eqnarray}
S_0&=&|E_+|^2+|E_-|^2=2 \exp\left(-\Lambda_+\right)\cosh\Lambda_-\, ,\nonumber\\
S_1&=&|E_+|^2-|E_-|^2=2 \exp\left(-\Lambda_+\right)\sinh\Lambda_-\, ,\nonumber\\
S_2&=& E_+E_-^*+E_+^*E_-=2 \exp\left(-\Lambda_+\right)\cos\delta\, ,\nonumber\\
S_3&=&i(E_+E_-^*-E_+^*E_-)=2 \exp\left(-\Lambda_+\right)\sin\delta\, ,
\label{stokesparametersy}
\end{eqnarray}
 where
\begin{eqnarray}
\Lambda_\pm&=&\frac{\omega}{c}\int \left(\pm \mbox{Im}\left[\chi_+\right]+\mbox{Im}\left[\chi_-\right] \right)dz\, ,\nonumber\\
\delta&=&\frac{\omega}{c}\int \left(\mbox{Re}\left[\chi_-\right]-\mbox{Re}\left[\chi_+\right] \right)dz\, .
\end{eqnarray}
These parameters are general, and they can be used to characterize any kind of supersymmetric polarized electromagnetic wave.  

\section{Exact supersymmetric solutions}
Eq.~\eqref{eqpolartioem} have a nonlinear behavior; however, some straightforward solutions can be found. In this Section, we propose some concrete $\zeta(z)$ functions, for Eq.\eqref{definedw}, such that Eq.~\eqref{eqpolartioem} can be solved exactly.\\

\subsection{Case for $\zeta(z)=\alpha z$}
\begin{figure}[H]
\centering
\subfloat[$\alpha=1$]
{\includegraphics[width=0.4\textwidth]{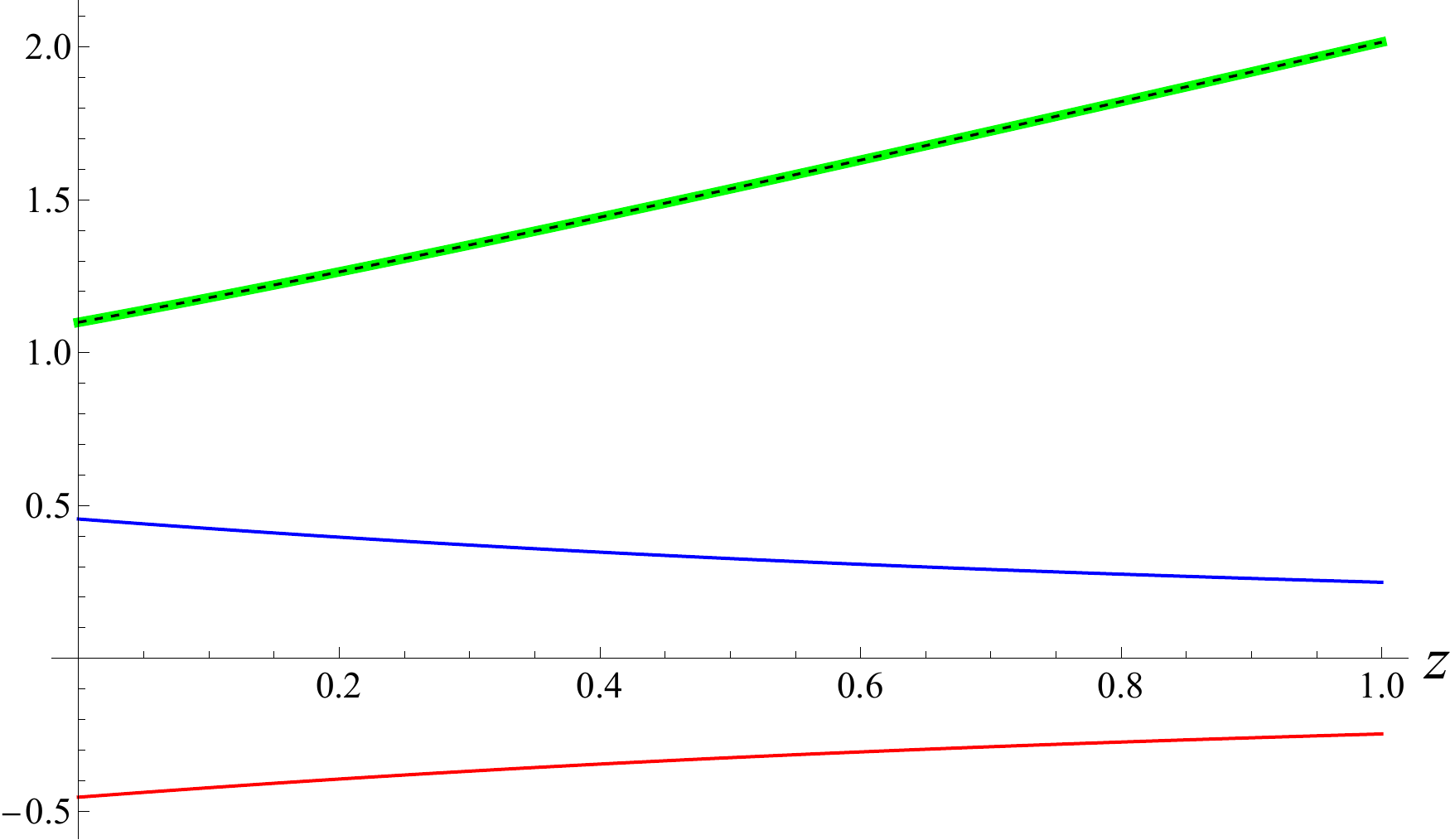}} \quad
\subfloat
{\includegraphics[width=0.1\textwidth]{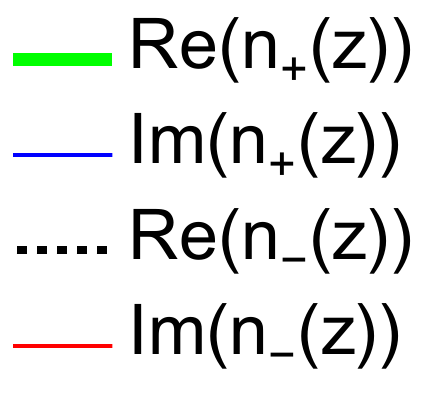}} \quad
\subfloat[$\alpha=1+i$]
{\includegraphics[width=0.4\textwidth]{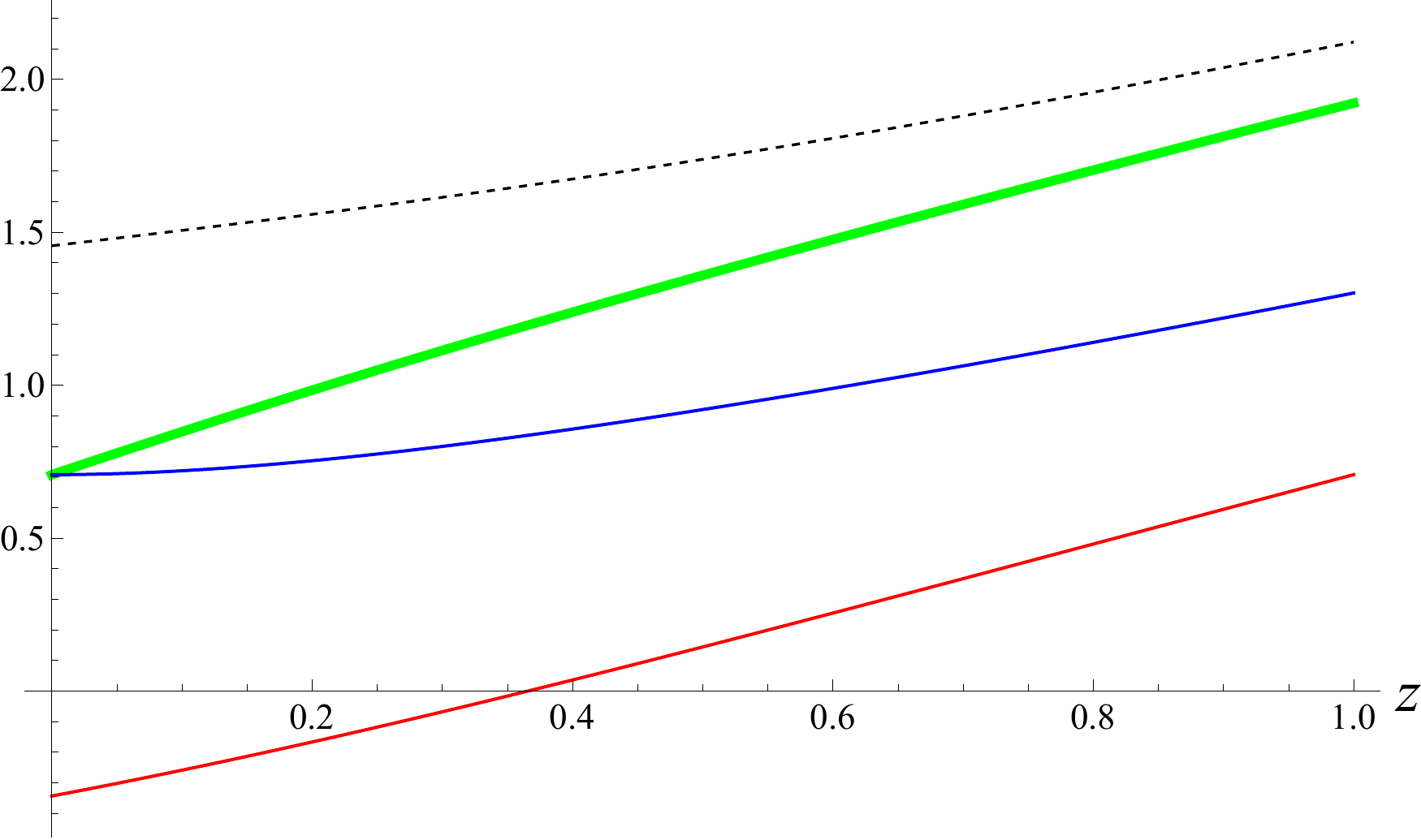}} 
\caption{The real and imaginary parts of $n_+(z)$ and $n_-(z)$ as functions of $z$, for $\alpha=1$ and $\alpha=1+i$, with $\omega=1$ and $c=1$.}\label{fig5}
\end{figure}

In this case, where $\alpha$ is a complex constant, we obtain the refractive indices from Eqs.~\eqref{permipluminus} and \eqref{definedw}, which are
\begin{equation}
n_\pm(z)=\sqrt[4]{\alpha^2 \frac{c^2}{\omega^2}+(1+\alpha z)^4}
\left[ \cos \left(\frac{1}{2} \arg \left(\omega  (\alpha  z+1)^2\pm i \alpha  c\right)\right)
+i \sin \left(\frac{1}{2} \arg \left(\omega  (\alpha  z+1)^2\pm i \alpha  c\right)\right) \right],
\end{equation}
In Fig.~\ref{fig5}, we depict the real and imaginary parts of $n_+(z)$ and $n_-(z)$ as functions of $z$, for $\alpha=1$, and $\alpha=1+i$.
Note that the imaginary part is non-zero for both indices as needed to get  supersymmetry. Also, the variation of the refractive indices can be tuned according to the needs, by varying the $\alpha$ parameter. 
In this case, Eq.~\eqref{eqpolartioem} can be solved exactly; the general solution is
\begin{equation}
\chi_+(z)=-2-\alpha z
+\frac{2 i \sqrt{\alpha } \; c \;
\exp \left[\frac{i}{\alpha}\frac{\omega}{c}(\alpha  z+1)^2\right]}
{(-1)^{3/4} \sqrt{\pi \omega c} \; 
\text{erfi}\left(\sqrt[4]{-1} \sqrt{\frac{\omega}{\alpha c}}(\alpha  z+1)\right)
-2 \beta_1 \sqrt{\alpha }  \omega  
\exp\left(\frac{i}{\alpha}\frac{\omega}{c}\right)},
\end{equation}
where $\beta_1$ is an arbitrary integration constant and $\text{erfi}(x)$ is the imaginary error function.
Finally, using Eq.~\eqref{formelectromagneticpolariz}, we can calculate both polarizations of the electric fields, getting
\begin{eqnarray}
E_-(z)&=&E_{0}\exp\left[ i \frac{\omega}{c} \left(z+\frac{\alpha}{2} z^2+\beta_2 \right) \right], \nonumber\\
E_+(z)&=&E_{0} \exp\left[ i \frac{\omega}{c} \left(z+\frac{\alpha}{2} z^2+f(z) \right) \right], 
\end{eqnarray}
being $\beta_2$ another arbitrary integration constant, and
\begin{equation}
f(z)=\gamma-\frac{\alpha  z^2}{2}-2 z
-i\frac{c}{\omega } 
\ln \left[-(-1)^{3/4} \sqrt{\pi c } \; \text{erfi}\left(\sqrt[4]{-1} \sqrt{\frac{\omega}{\alpha c}}(\alpha  z+1)\right)+2 \beta_1 \sqrt{\alpha \omega} \;
\exp\left(\frac{i}{\alpha}\frac{\omega}{c}\right)\right],
\end{equation}
with $\gamma$ yet another integration constant.
In the next figure, we show the intensity of the two polarization's of the electric field for the same values of the parameter $\alpha$,
\begin{figure}[H]
        \centering
	\subfloat[$\alpha=1$]
	{\includegraphics[width=0.48\textwidth]{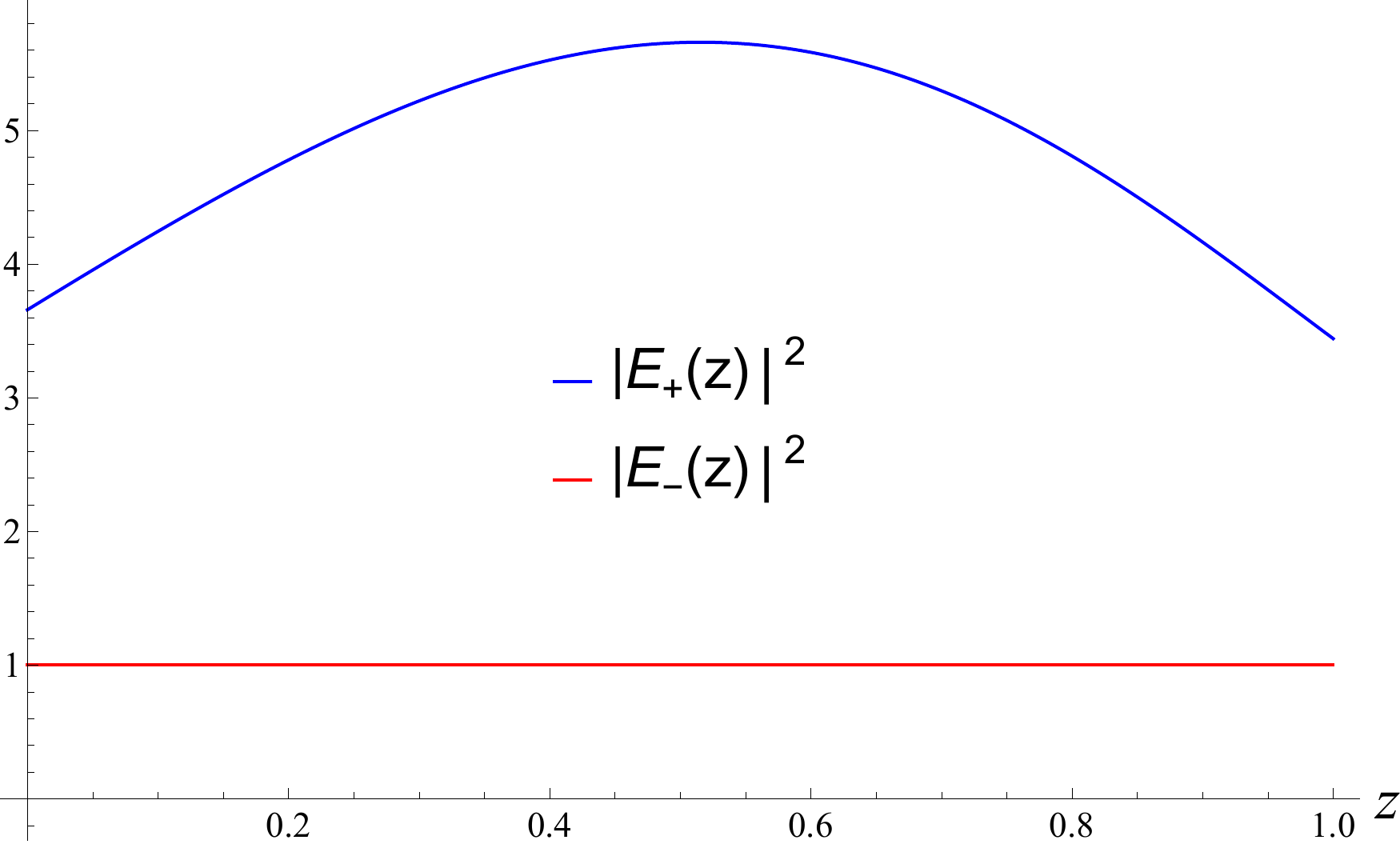}}\hfill
	\subfloat[$\alpha=1+i$]
	{\includegraphics[width=0.48\textwidth]{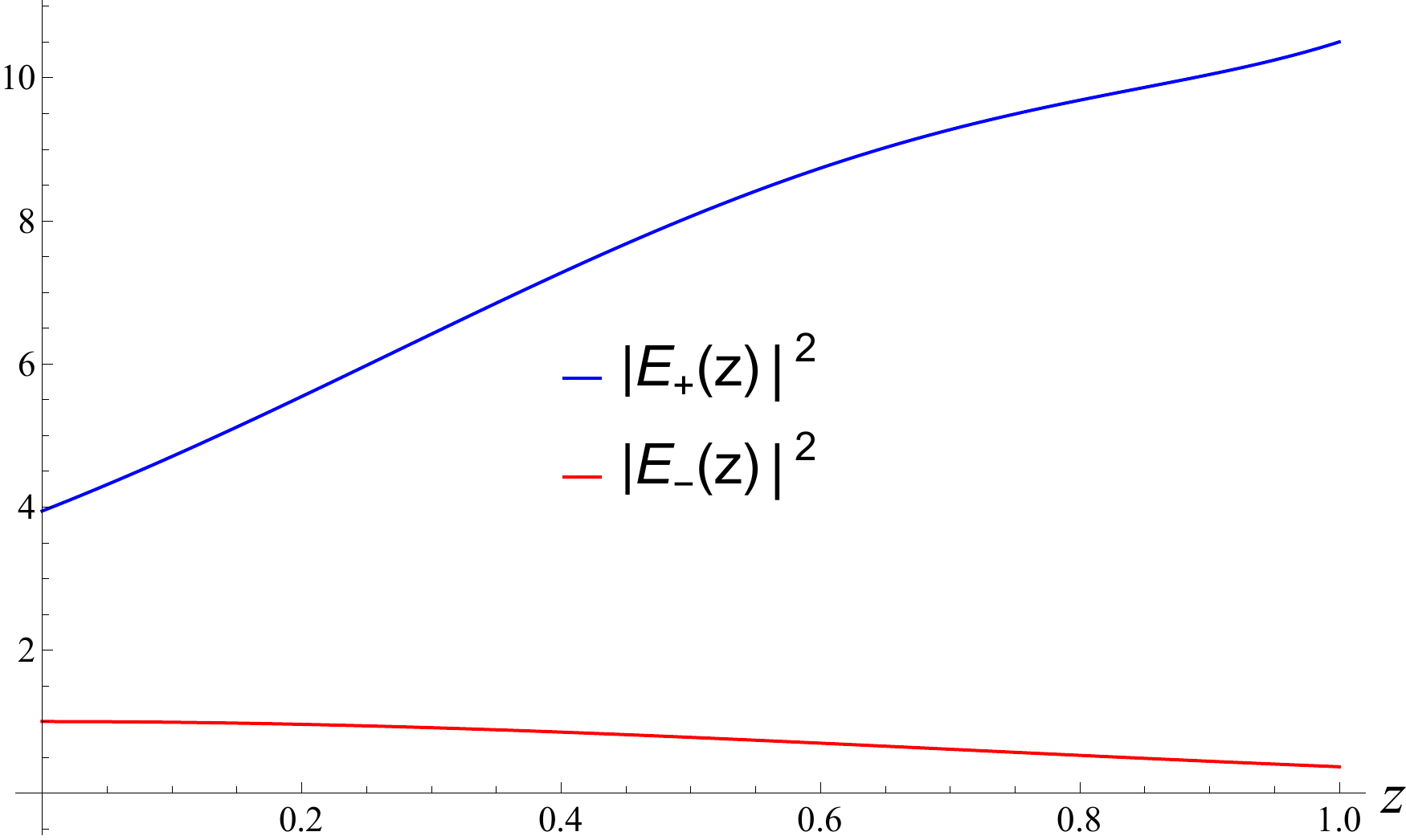}} 
	\caption{The intensity of the two polarizations of the electric field with the previous two values of $\alpha$. The parameters used are $E_{0}=1$, $c=1$, $\omega=1$, $\beta_1=0$, $\beta_2=0$, and $\gamma=0$}
\end{figure}

\subsection{Case for $\zeta(z)=\alpha \exp\left( \beta z\right)$}
\begin{figure}[H]
    \centering
	\subfloat[The refractive indices for $\zeta(z)=\alpha \exp\left( \beta z\right)$, with $\alpha=0.1$, $\beta=-1.0$, $\omega=1.0$, and $c=1.0$\label{fig3a}]
	{\includegraphics[width=0.48\textwidth]{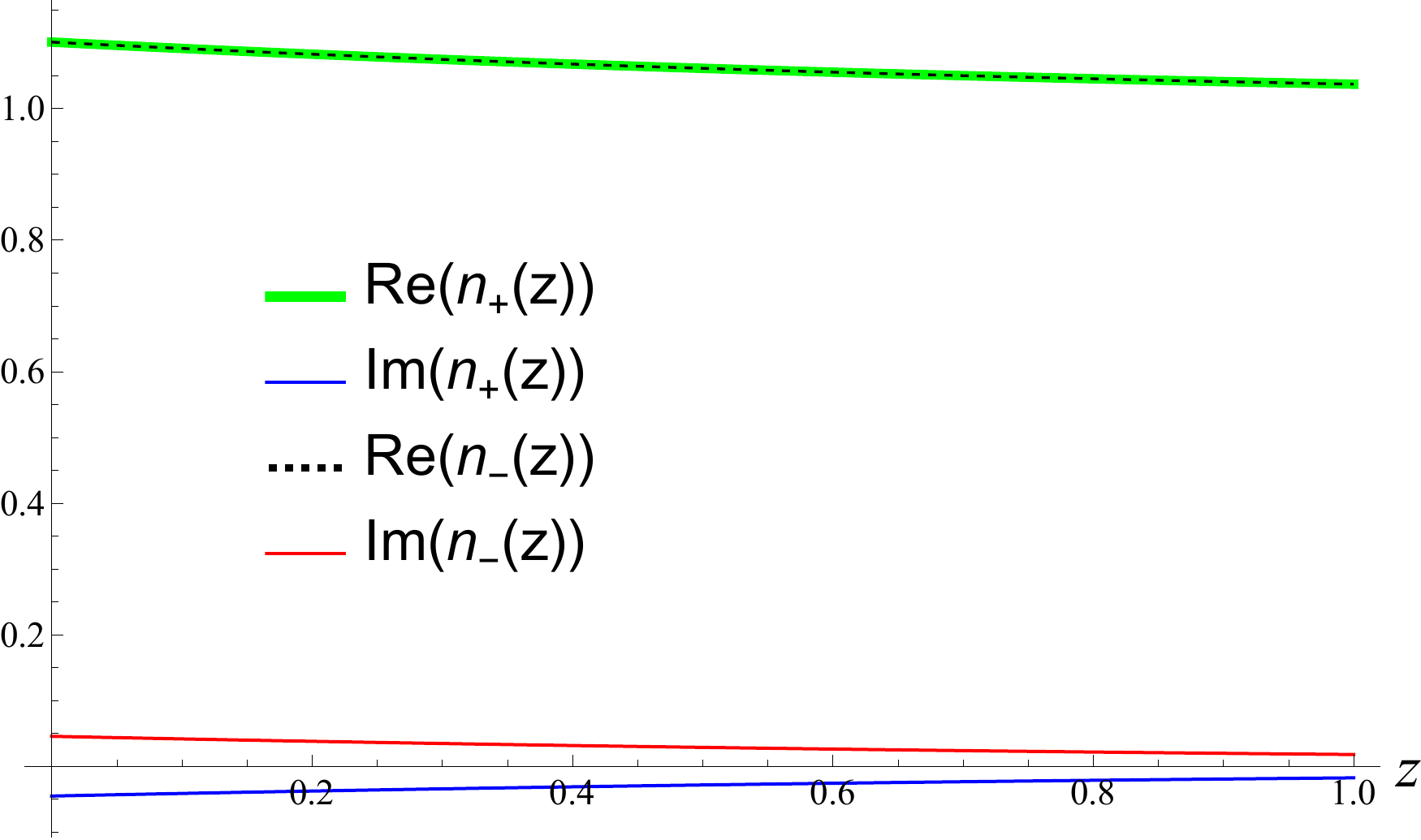}}\hfill
	\subfloat[The intensity of the two polarizations of the electric field for $\alpha=0.1$, and $\beta=-1.0$. The parameters used are $E_{0 +}=1$, $E_{0 -}=1$, $c=1$, $\omega=1$, $\delta=0$, $\eta=0$.\label{fig3b}]
	{\includegraphics[width=0.48\textwidth]{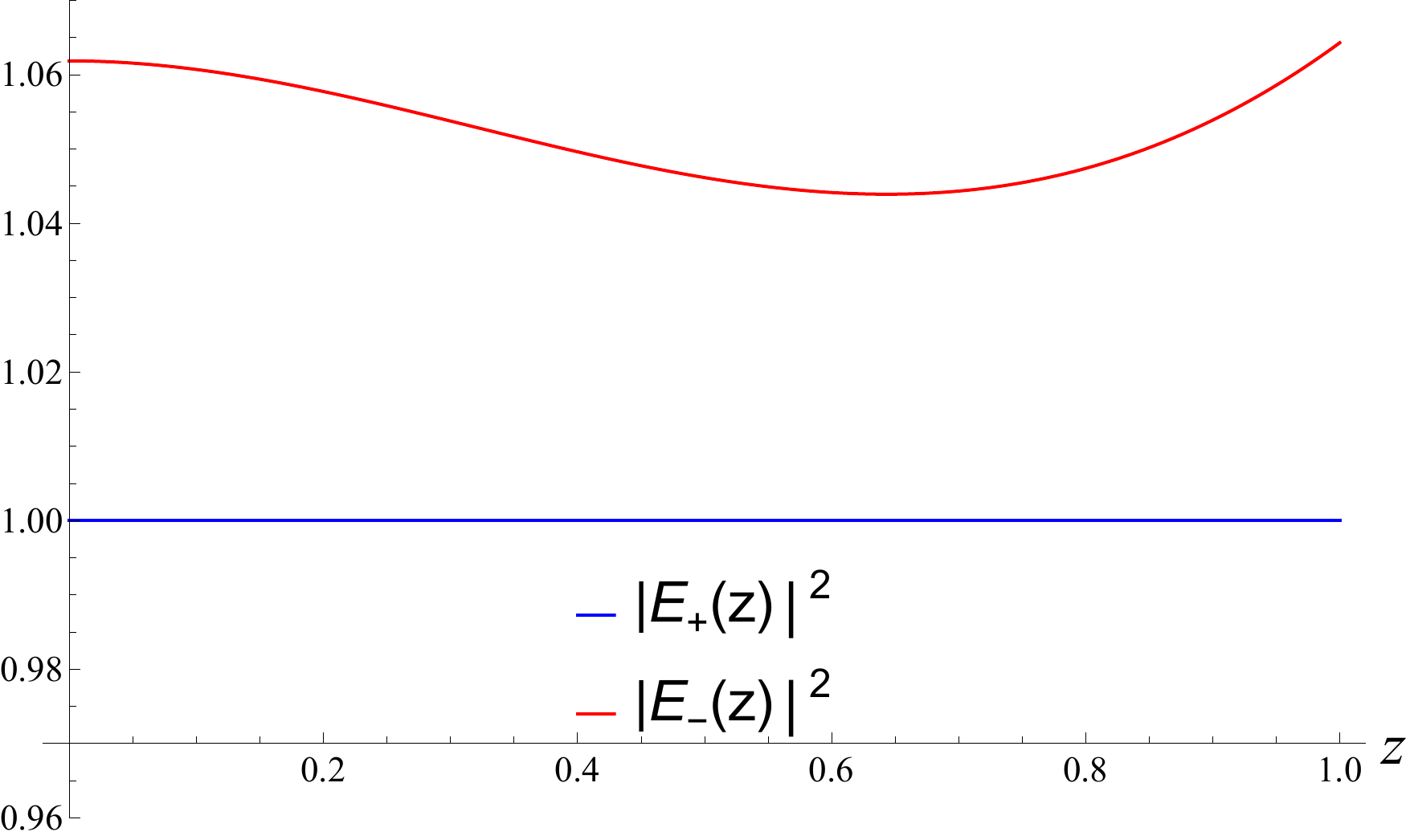}} 
	\caption{}
	\label{fig3}
\end{figure}

For this case, when  $\alpha$ and $\beta$ are complex constants, we obtain the refractive indices,
\begin{eqnarray}
n_\pm(z)&=&\frac{\sqrt[4]{\alpha ^2 \beta ^2 c^2 \omega ^2 e^{2 \beta  z}+\left(\omega +\alpha  \omega  e^{\beta  z}\right)^4}}{\omega }
\left[ 
\cos \left(\frac{1}{2} \arg \left(\omega  \left(\alpha  e^{\beta  z}+1\right)^2\pm i \alpha  \beta  c e^{\beta  z}\right)\right)\right.\nonumber\\
&&\qquad\qquad\qquad\qquad\qquad \left.+i \sin \left(\frac{1}{2} \arg \left(\omega  \left(\alpha  e^{\beta  z}+1\right)^2\pm i \alpha  \beta  c e^{\beta  z}\right)\right)\right],
\end{eqnarray}
The function $\zeta(z)=\alpha \exp(\beta z)$, and the refractive indices derived from it, just above, offer many possibilities for behavior when the parameters $\alpha$ and $\beta$ are complex numbers; however, as we will see later, in order to calculate the electric field, we must integrate over certain combinations of them, and this will restrict us to considering that $\alpha>0$, $\beta<0$, and that both are small. Therefore, in Fig.~\ref{fig3a}, we consider the case when $\alpha=0.1$, and $\beta=-1.0$; in that figure, we depict the real and imaginary parts of $n_+(z)$ and $n_-(z)$ as functions of $z$. Again, the imaginary part is non-zero for both indices, and the variation of the refractive indices can be tuned according to the needs, by varying the $\alpha$ and $\beta$ parameters.
In this case, Eq.~\eqref{eqpolartioem} can be solved exactly; the  general solution is
\begin{align}\label{0260}
\chi_+(z)=&-\frac{\alpha  e^{\beta  z} 2^{\frac{2 i \omega }{\beta  c}} \left(\frac{i \alpha  \omega  e^{\beta  z}}{\beta  c}\right)^{\frac{2 i \omega }{\beta  c}} L_{-\frac{2 i \omega }{c \beta }}^{\frac{2 i \omega }{c \beta }}\left(\frac{2 i e^{z \beta } \alpha  \omega }{c \beta }\right)}{\gamma_1+2^{\frac{2 i \omega }{\beta  c}} \left(\frac{i \alpha  \omega  e^{\beta  z}}{\beta  c}\right)^{\frac{2 i \omega }{\beta  c}} L_{-\frac{2 i \omega }{c \beta }}^{\frac{2 i \omega }{c \beta }}\left(\frac{2 i e^{z \beta } \alpha  \omega }{c \beta }\right)}
-\frac{\alpha  e^{\beta  z} 2^{1+\frac{2 i \omega }{\beta  c}} \left(\frac{i \alpha  \omega  e^{\beta  z}}{\beta  c}\right)^{\frac{2 i \omega }{\beta  c}} L_{-\frac{2 i \omega }{c \beta }-1}^{\frac{2 i \omega }{c \beta }+1}\left(\frac{2 i e^{z \beta } \alpha  \omega }{c \beta }\right)}{\gamma_1+2^{\frac{2 i \omega }{\beta  c}} \left(\frac{i \alpha  \omega  e^{\beta  z}}{\beta  c}\right)^{\frac{2 i \omega }{\beta  c}} L_{-\frac{2 i \omega }{c \beta }}^{\frac{2 i \omega }{c \beta }}\left(\frac{2 i e^{z \beta } \alpha  \omega }{c \beta }\right)}
\nonumber \\ &
-\frac{2 \gamma_1}{\gamma_1+2^{\frac{2 i \omega }{\beta  c}} \left(\frac{i \alpha  \omega  e^{\beta  z}}{\beta  c}\right)^{\frac{2 i \omega }{\beta  c}} L_{-\frac{2 i \omega }{c \beta }}^{\frac{2 i \omega }{c \beta }}\left(\frac{2 i e^{z \beta } \alpha  \omega }{c \beta }\right)}
-\frac{\alpha  \gamma_1 e^{\beta  z}}{\gamma_1+2^{\frac{2 i \omega }{\beta  c}} \left(\frac{i \alpha  \omega  e^{\beta  z}}{\beta  c}\right)^{\frac{2 i \omega }{\beta  c}} L_{-\frac{2 i \omega }{c \beta }}^{\frac{2 i \omega }{c \beta }}\left(\frac{2 i e^{z \beta } \alpha  \omega }{c \beta }\right)}
\end{align}
where $\gamma_1$ is an arbitrary integration constant and $L_\nu^\mu(\xi)$ are the Laguerre generalized polynomials.
As can be seen in Eq.~\eqref{formelectromagneticpolariz}, to calculate the polarization of the electric field $E_+(z)$, we need to integrate $\chi_+(z)$ with respect to $z$, and with expression \eqref{0260} that it is difficult; hence, we impose the initial condition $\chi_+(0)=0$, and we assume  that $\alpha>0$, $\beta<0$, and that both are small ($\left|\alpha \right| \ll 1$ and $\left|\beta \right| \ll 1$). With this condition and those assumptions, we develop in Taylor series the solution \eqref{0260}, to finally get
\begin{equation}
\chi_+(z)=-\frac{\alpha  e^{-\frac{2 i \omega  z}{c}}}{\beta  c+2 i \omega }
\left[  -\beta c+e^{\frac{2 i \omega  z}{c}} (\beta  c (\beta  z+1)+2 i \omega  (\beta  z-1))+2 i \omega \right], 
\end{equation}
which can be integrated with respect to $z$.
Finally, using Eq.~\eqref{formelectromagneticpolariz}, we can calculated both polarizations of the electric fields, getting
\begin{equation}
E_\pm(z)=E_{0} \exp\left\lbrace  i \frac{\omega}{c} \left[z+f_\pm(z) \right] \right\rbrace , 
\end{equation}
where
\begin{equation}
f_+(z)=\delta -\frac{\alpha }{2 (\beta  c+2 i \omega )}
 \left(\frac{c (-2 \omega +(-i) \beta  c) e^{-\frac{2 i \omega  z}{c}}}{\omega }+\beta  c z (\beta  z+2)+2 i \omega  z (\beta  z-2)\right),
\end{equation}
being $\delta$ an integration constant,and
\begin{equation}
f_-(z)=\eta+\frac{\alpha  e^{\beta  z}}{\beta },
\end{equation}
with $\eta$ yet another integration constant. In Fig.~\ref{fig3b}, we show the intensity of the two polarization's of the electric field for the same values of the parameter $\alpha$,

\subsection{Case for $\zeta(z)=-1+\sqrt{{\alpha c}/{\omega}}\, 
\tanh \left(\sqrt{{\alpha \omega}/{c}}\, (\beta c-i z)\right)$}
\begin{figure}[H]
    \centering
	\subfloat[The refractive indices for $\zeta(z)=-1+\sqrt{\frac{\alpha c}{\omega}}
	\tanh \left(\sqrt{\frac{\alpha \omega}{c}}(\beta c-i z)\right)$, when  $\alpha=1+i$ and $\beta=i$.\label{fig4a}]
	{\includegraphics[width=0.48\textwidth]{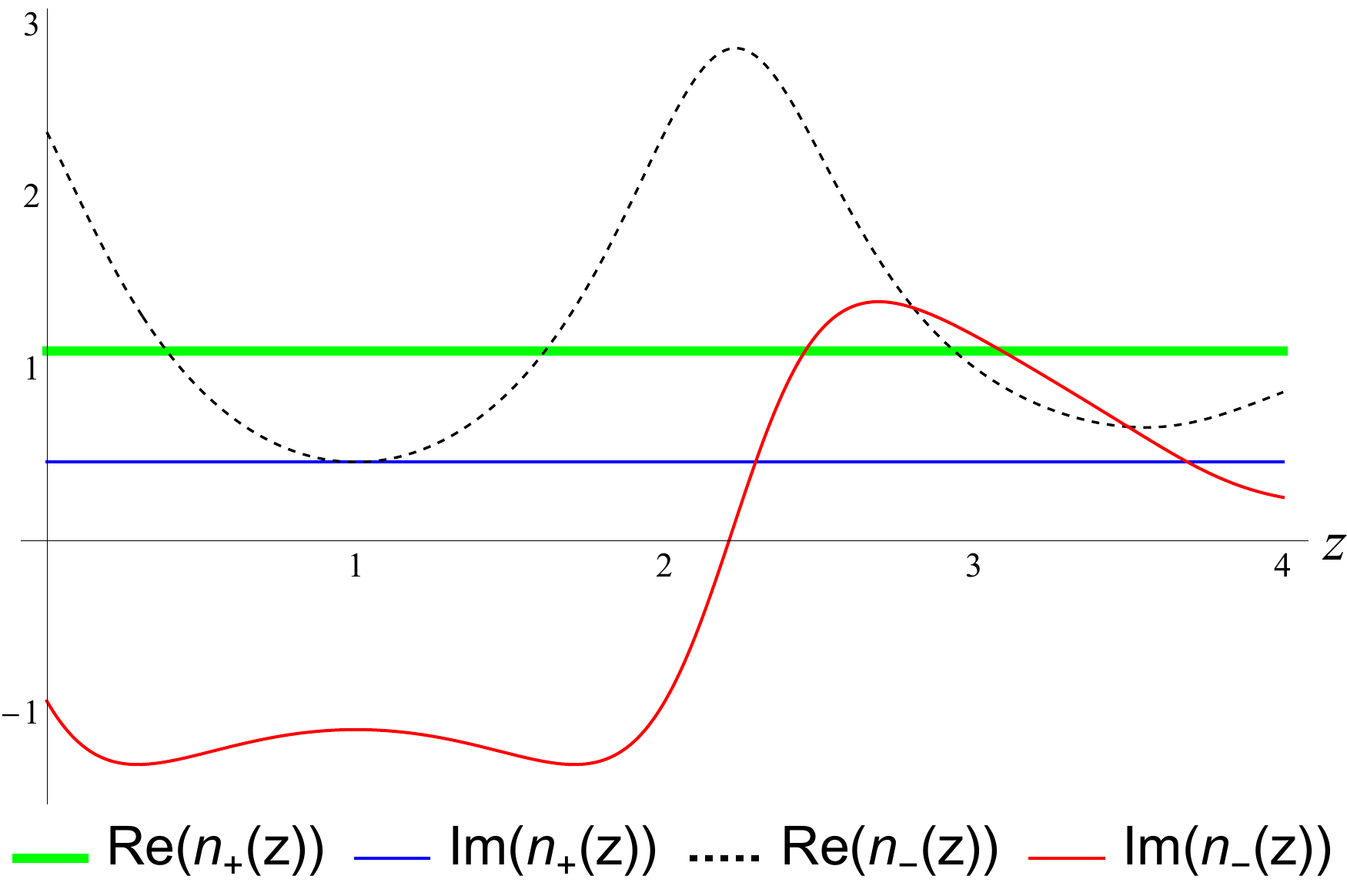}}\hfill
	\subfloat[The intensity of the two polarizations of the electric field for $\alpha=1+i$, and $\beta=i$. The parameters used are $E_{0 +}=1$, $E_{0 -}=1$, $c=1$, $\omega=1$, $\gamma=0$.\label{fig4b}]
	{\includegraphics[width=0.48\textwidth]{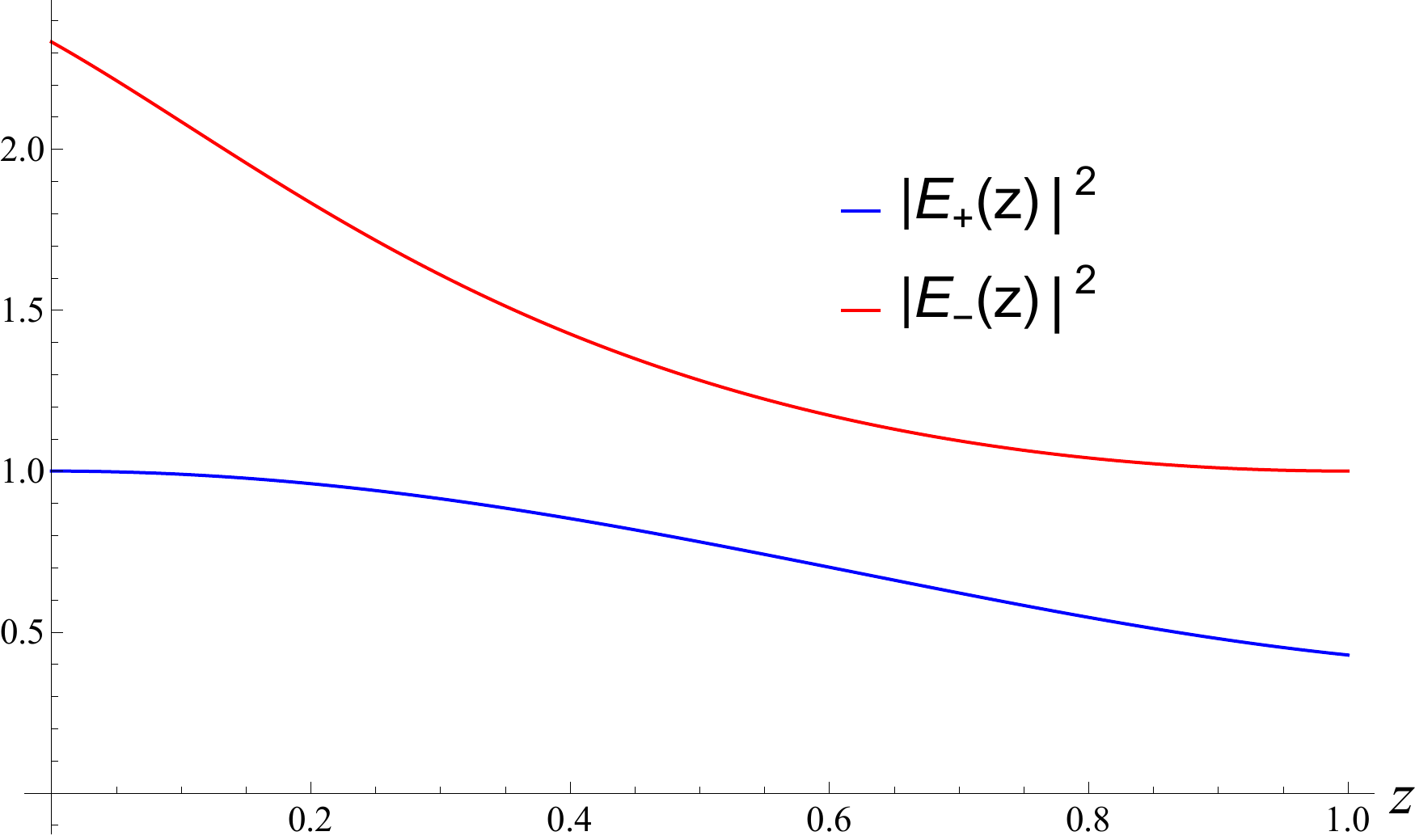}}
	\caption{}
	\label{fig4}
\end{figure}

We propose that
for $\alpha$ and $\beta$ being complex constants, then using Eqs. \eqref{permipluminus} and \eqref{definedw}, we obtain the refractive indices,
\begin{equation}
n_+(z)=\sqrt{\frac{\alpha c}{\omega }}
\left[ \cos \left(\frac{\arg (\alpha )}{2}\right)+i
\sin \left(\frac{\arg (\alpha )}{2}\right)\right] ,
\end{equation}
and
\begin{align}
n_-(z)=&\sqrt{\frac{\alpha  c}{\omega }} 
\left[ \cos \left(\frac{1}{2} \arg \left(2 \alpha -4 \alpha  \sec ^2\left(\sqrt{\frac{\alpha  \omega }{c}} (z+i \beta  c)\right)\right)\right)+
i \sin \left(\frac{1}{2} \arg \left(2 \alpha -4 \alpha  \sec ^2\left(\sqrt{\frac{\alpha  \omega }{c}} (z+i \beta  c)\right)\right)\right)\right] 
\nonumber \\ &
\times \frac{\sqrt[4]{\left(-3+\cos \left(2 e^{-\frac{1}{2} i \arg (\alpha )} \sqrt{\frac{\alpha  \omega }{c}} (z-i \beta  c)\right)\right) \left(-3+\cos \left(2 e^{\frac{1}{2} i \arg (\alpha )} \sqrt{\frac{\alpha  \omega }{c}} (z+i \beta  c)\right)\right)}}{\sqrt{\cos \left(2 \sqrt{\frac{\alpha  \omega }{c}} \left(\beta  c \sin \left(\frac{\arg (\alpha )}{2}\right)-z \cos \left(\frac{\arg (\alpha )}{2}\right)\right)\right)+\cosh \left(2 \sqrt{\frac{\alpha  \omega }{c}} \left(\beta  c \cos \left(\frac{\arg (\alpha )}{2}\right)+z \sin \left(\frac{\arg (\alpha )}{2}\right)\right)\right)}}
\end{align}
For $\alpha=1+i$, and $\beta=i$, we get the behavior depicted in Fig.~\ref{fig4a}. We can note that the refractive index $n_+(z)$ is independent of $z$. Also, both indices have imaginary part. In this case, Eq.~\eqref{eqpolartioem} can be solved exactly; the  solution is
\begin{equation}
\chi_+(z)=-1+\sqrt{\frac{\alpha c}{\omega}}
\tanh \left(\sqrt{\frac{\alpha \omega}{c}}(\beta c+i z)\right),
\end{equation}
where $\gamma$ is an arbitrary integration constant.
Finally, using Eq.~\eqref{formelectromagneticpolariz}, we can calculated both polarizations of the electric fields, getting
\begin{equation}
E_\pm(z)=E_{0} \exp\left\lbrace  i \frac{\omega}{c} \left[z+p_\pm(z) \right] \right\rbrace , 
\end{equation}
where
\begin{equation}
p_+(z)=-z-\frac{i c }{\omega }
\ln \left(\cos \left(\sqrt{\frac{\alpha \omega}{c}}(z-i c \gamma )\right)\right),
\end{equation}
being $\gamma$ an integration constant, and
\begin{equation}
p_-(z)=-z+\frac{i c }{\omega }
\ln \left(\cos \left(\sqrt{\frac{\alpha \omega}{c}}(z+i \beta  c)\right)\right).
\end{equation}
In Fig.~\ref{fig4b}, we show the intensity of the two polarization's of the electric field for the same values of the parameter $\alpha=1+i$, and $\beta=i$.

\section{Supersymetric cases for small correction to refractive indexes}
In this section, we will explore  cases in which \eqref{permipluminus} can be fulfilled in a general approximated way, when the refractive indexes have a small correction that allows to emerge the supersymmetric characteristics of the electromagnetic wave. In this case, we can argue that $\left|\zeta(z)\right| \ll 1$, for all $z$, and the anisotropic refractive indices acquire the form
\begin{align}
n_\pm(z)&=  1+\zeta\pm i\frac{c}{2 \omega}\frac{d\zeta}{dz}\, .
\label{refractionsmall}
\end{align}
Because of the complex nature of these refractive indexes in this supersymmetric theory, it is implied that as the wave propagates, its amplitude suffer decaying or amplification by the medium. 
Therefore,  a medium with the  refractive indexes \eqref{refractionsmall} for a small anisotropy, induces different polarizations that are completely determined by Eqs.~\eqref{eqpolartioemmenos} and \eqref{eqpolartioem}, solved as $\chi_-=\zeta$, and 
\begin{eqnarray}
  \chi_+= e^{-2i\omega z/c}\int dz \left(-\frac{d\zeta}{dz}+\frac{2i\omega}{c}\zeta\right)e^{2i\omega  z/c}\, ,
  \label{soluctonschi}
\end{eqnarray}
where we have neglected the $\chi_+^2$ term in Eq.~\eqref{eqpolartioem}, as it is assumed to be a small correction. In this way, we are assuming that $\chi_+$ depends linearly on $\zeta$ and its derivatives.   

\subsection{Simplest real function $\zeta=z/L_1+z^2/L_2^2$}
Let us assume this simplest possible real choice for the perturbation function, where $z\ll L_1,L_2$ in order to maintain the approximation. In such case, the refractive indexes \eqref{refractionsmall} are 
\begin{align}
n_\pm(z)&=  1+\frac{z}{L_1}+\frac{z^2}{L_2^2}\pm i\frac{c}{2 \omega L_1}\left(1+\frac{2L_1 z}{L_2^2}\right)\, ,
\label{refractionsmallsmall}
\end{align}
obtaining quadratic indices medium, with imaginary parts that can be chosen as small as desired by an appropriated choice of $L_1$, $L_2$ or $\omega$.

In this case $\chi_-=\zeta$, and $\chi_+=\zeta-c^2/(L_2^2\omega^2)+ic/(L_1\omega)+2 i c z/(L_2^2\omega)$. Thereby, the electromagnetic polarizations behave supersymmetrically in the medium given by \eqref{refractionsmallsmall}, and can been found by solving Eqs.~\eqref{soluctonschi},
\begin{eqnarray}
E_+(z)&=&\exp\left(-\zeta \right) \exp\left[i \frac{\omega}{c} \left(z\left(1-\frac{c^2}{\omega^2 L_2^2}\right)+\frac{z^2}{2L_1}+\frac{z^3}{3L_2^2}\right)\right]\, ,\nonumber\\
E_-(z)&=& \exp\left[i \frac{\omega}{c} \left(z+\frac{z^2}{2L_1}+\frac{z^3}{3L_2^2}\right)\right]\, .
\end{eqnarray}
For this case, only the $E_+$ polarization presents attenuation as it propagates. However, this is not a large effect due to our assumptions. This can also be seen by the Stokes parameters \eqref{stokesparametersy}. We obtain that
$\Lambda_\pm=\pm \zeta$, and $\delta=c z/(L_2^2\omega)$. Therefore, the intensity of the polarized wave decays as $\exp(-\zeta)$, as it is seen through $S_0$. Also, as $S_1<0$, the electromagnetic has a tendency to be vertical linear polarized. In case that $z/L_2\lll L_2\omega/c$, then $\delta\approx 0$, and the wave becomes linear polarized.

\subsection{Simplest imaginary function $\zeta=i z/L$ }

Another interesting example for the case of small corrections, when $z\ll L$, takes place for this purely imaginary case. In this conditions, the refractive indices become
\begin{align}
n_\pm(z)&=  1\mp \frac{c}{2 \omega L}+i\frac{z}{L}\, ,
\label{refractionsmallsmall2}
\end{align}
which is the opposite behavior compared to refractive indices \eqref{refractionsmallsmall}. The supersymmetric polarizations of the electromagnetic waves propagate in this medium as
\begin{eqnarray}
E_+(z)
&=&\exp\left(-\frac{\omega z^2}{2cL} \right) \exp\left[i \frac{\omega}{c} z\left(1-\frac{c}{\omega L}\right)\right]\, ,\nonumber\\
E_-(z)&=&\exp\left(-\frac{\omega z^2}{2cL} \right) \exp\left[i \frac{\omega}{c} z\right]\, ,
\end{eqnarray}
as $\chi_-=\zeta$, and $\chi_+=-c/(L\omega)+\zeta$. 
These two superpartner polarized waves present the same quadratic decaying along its propagation in the longitudinal direction, and they have a phase shift of the order of $\sim z/L$. These features are, again, seen through the Stokes parameters \eqref{stokesparametersy}. In this case, we calculate that $\Lambda_+=\omega z^2/(c L)$, $\Lambda_-=0$, and $\delta=z/L$. These results imply that the intensity ($S_0$) decays as $\exp(-\omega z^2/cL)$. Besides, this wave does not have a preferred polarization, as
$S_1=0$.

\section{Discussions}

This theory, with its corresponding solutions, allows us to establish that electromagnetic waves can classically behave in an analog fashion to a supersymmetric quantum mechanical system. This is achieved through the superpartnership of the wave's polarizations. This is the first time that this supersymmetric realization is proposed for electromagnetic wavepackets, where it is taken advantage of the spatial extension of the field, and its interaction with a medium, to construct this special behavior of light.

The above ideas open up the exciting possibility of studying supersymmetry at an optical level using anisotropic optical materials. The difference of the current proposal with others, lies in the study of the entire dynamics of the polarized electromagnetic wave to investigate its supersymmetric features, rather than just its behavior in the eikonal (geometrical) limit that models light as rays. In the latter case, two different media are required to model supersymmetric behavior in scalar light, whereas for polarized light, as we have seen, only one anisotropic medium is needed. Thus, we have extended previous notions of supersymmetry in optical systems.

We have proven that this supersymmetric behavior can be analytical studied. For this, we have presented exact solutions for media in which this supersymmetric behavior of light can be tested. Furthermore, it can also be demonstrated this effect in simple media by using small linear (complex) corrections to their anisotropic refractive indices. Therefore, checking the supersymmetric form of an electromagnetic wave should be straightforward in optical systems.
In this sense, it is important to highlight that our proposal can be experimentally tested in various anisotropic media, such as birefringent crystals or liquid crystals. In these materials, the refractive indices can be controlled and varied, allowing for the observation of the predicted supersymmetric behavior in polarized electromagnetic waves. Additionally, with the use of modern optical techniques, such as polarization-modulation  and phase-sensitive detection (see for example Refs.~\cite{wolf22,FAModine,FAModine2,Marguerre,AZGenack,KJVeenstra}), the measurement of the superpartner behavior of the polarizations can be achieved with high accuracy. These experimental demonstrations would provide strong evidence for the classical analog of supersymmetry in electromagnetic waves, joining previous efforts \cite{MatthiasHeinrich,kwolf,carlos}, and 
opening new possibilities for the exploration of fundamental physics and its applications in optics.

\begin{acknowledgements}
FAA thanks to FONDECYT grant No. 1230094 that supported this work.
 \end{acknowledgements}

\end{document}